\documentclass[preprint,12pt,number]{elsarticle}
\usepackage{float}
\usepackage{tikz}
\usepackage{amsmath}
\usepackage{csquotes}
\usepackage{hyperref}
\usepackage{graphicx}
\usepackage{subcaption}
\journal{Computer Communications}

\begin{document}
\begin{frontmatter}

\title{Path Dynamics in a Deployed Path-Aware Network:\\A Measurement Study of SCIONLab}

\author[TRU,Goethe]{Lars Herschbach\corref{cor1}\fnref{fn1}}
\ead{Lars.Herschbach@t-online.de}

\author[TRU,Cesi]{Damien Rossi\fnref{fn2}}
\ead{damien.rossi@viacesi.fr}

\author[TRU]{Sina Keshvadi}
\ead{skeshvadi@tru.ca}

\cortext[cor1]{Corresponding author.}
\fntext[fn1]{Present address: Goethe University, Frankfurt, Germany.}
\fntext[fn2]{Present address: Cesi School of Engineering, Toulouse, France.}

\address[TRU]{Thompson Rivers University, BC, Canada}
\address[Goethe]{Goethe University, Frankfurt, Germany}
\address[Cesi]{Cesi School of Engineering, Toulouse, France}

\begin{abstract}
    Path-aware networks promise enhanced performance and resilience through multipath transport, but a lack of empirical data on their real-world dynamics hinders the design of effective protocols. This paper presents a longitudinal measurement study of the SCION architecture on the global SCIONLab testbed, characterizing the path stability, diversity, and performance crucial for protocols like Multipath QUIC (MPQUIC). Our measurements reveal a dynamic environment, with significant control-plane churn and short path lifetimes in parts of the testbed. We identify and characterize path discrepancy, a phenomenon where routing policies create asymmetric path availability between endpoints. Furthermore, we observe a performance trade-off where concurrent multipath transmissions can improve aggregate throughput but may degrade the latency and reliability of individual paths. These findings demonstrate that protocols such as MPQUIC should explicitly account for high churn and path asymmetry, challenging common assumptions in multipath protocol design.

\end{abstract}


\end{frontmatter}

\section{Introduction}~\label{sec::introduction}
    The Internet's ongoing evolution is driven by relentless demands for greater security, resilience, and performance. Yet its inter-domain routing system, built on BGP, offers little control or visibility to endpoints and remains prone to disruption~\cite{giaconia2025crunching}. Path changes are opaque, failures can take minutes to recover from, and malicious or accidental route announcements can divert traffic through insecure or suboptimal paths~\cite{butler2009survey}. These limitations constrain the design of advanced transport protocols and complicate efforts to build a secure, scalable, performance-aware, and policy-compliant network.
    
    To address these challenges, several next-generation network architectures have been proposed~\cite{clark2018designing}, each rethinking the relationship between endpoints and the underlying network. One such architecture is SCION (Scalability, Control, and Isolation on Next-Generation Networks)~\cite{zhang2011scion}, a path-aware Internet architecture that has moved beyond the research lab into real-world deployments across academia, industry, and government networks~\cite{anapaya2025}. SCION gives endpoints explicit control over the exact sequence of Autonomous Systems (ASes) their traffic traverses, enabling them to discover and select from multiple, diverse, and independent end-to-end paths.
    
    This capability aligns with growing interest in multipath transport protocols. Multipath TCP~\cite{rfc8684} has shown the benefits of using heterogeneous links (e.g., Wi-Fi and LTE) in parallel. The forthcoming Multipath QUIC (MPQUIC)~\cite{de2017multipath}, under standardization in the IETF~\cite{ietf-quic-multipath-15}, extends this concept to QUIC's flexible, encrypted transport layer. Multipath transport can enhance throughput, resilience, and load balancing by using multiple network paths simultaneously. However, today's multipath protocols have largely been designed and evaluated in environments with only a handful of alternative paths (e.g., IPv4/IPv6 or Wi-Fi/Cellular). In contrast, a path-aware architecture like SCION can expose a much richer set of paths, potentially with far greater diversity in latency, bandwidth, and reliability~\cite{krahenbuhl2024glids}.
    
    These observations raise important questions: Are SCION-exposed paths stable enough for transport protocols to use them effectively? Are paths symmetric between endpoints? Does concurrent use of many paths always improve performance compared to a single, well-chosen path?
    These unknowns are critical. 
    Answers to these questions are critical because MPQUIC's path selection, packet scheduling, and congestion control depend not only on the availability of multiple paths but also on their dynamic characteristics. SCION's control plane continuously creates and refreshes path segments through beaconing, making its routing environment inherently more dynamic than the relatively static paths in today's Internet~\cite{krahenbuhl2021deployment}. Without empirical data, protocol design risks relying on assumptions derived from simulations or traditional Internet deployments that may not hold in SCION.

    In this work, we address these questions through a four-week longitudinal measurement study on the SCIONLab testbed~\cite{kwon2020scionlab}. Our findings reveal that the path environment is far more dynamic than in the traditional Internet, exhibiting significant path churn that impacts stability. We identify and characterize \textit{path discrepancy}, a novel phenomenon where routing policies create asymmetric path availability between endpoints. 
    Critically, our measurements reveal a direct and unexpected performance trade-off. Naive concurrent multipath transmission can degrade the latency and reliability of each constituent path, challenging the common assumption that more is always better. 
    These results provide crucial, data-driven insights for the design of the next generation of multipath transport protocols.
    
    Our main contributions are:
    \begin{enumerate}
        \item An empirical characterization of SCIONLab's path dynamics, revealing significant control-plane churn with average path lifetimes as short as 8.6 hours.
        
        \item The identification and analysis of two key phenomena affecting multipath transport: 1)~ path discrepancy, where routing policies create asymmetric path availability, and 2)~a fundamental performance trade-off where concurrent usage improves aggregate throughput but degrades per-path latency and reliability.
        
        \item Design implications for multipath protocols like MPQUIC derived from these findings, with specific recommendations for schedulers to handle high churn, path asymmetry, and performance trade-offs.
        
        \item The development and public release of a measurement suite for SCION, enabling reproducible research on the dynamics of path-aware networks.
    \end{enumerate}
        
    The remainder of this paper is structured as follows. Section~\ref{sec::backgroundAndRelatedwork} discusses SCION and related multipath transport work. Section~\ref{sec::methodology} details our measurement methodology. Section~\ref{sec::result} presents the results, Section~\ref{sec::discussion} discusses implications, and Section~\ref{sec::conclusion} concludes.

\section{Background and Related Works}~\label{sec::backgroundAndRelatedwork}
    This section provides the necessary background to understand the contributions of this work. It begins with an overview of the SCION architecture, focusing on the dynamic path environment created by its beaconing process, and then situates our research within the existing literature.

\subsection{Background}~\label{subsec::background}
    SCION~\cite{zhang2011scion} is a clean-slate Internet architecture designed for high availability, security, and fine-grained path control. A central innovation is path awareness. End-hosts can explicitly choose the sequence of Autonomous Systems (ASes) their packets traverse. To improve scalability and fault isolation, SCION organizes ASes into Isolation Domains (ISDs) with their own trust roots and core ASes. Routing issues in one ISD are less likely to propagate beyond its boundaries, while inter-ISD connectivity is handled by core ASes.

    SCION’s design cleanly separates the control and data planes. Path discovery occurs in the control plane through beaconing, a continuous process in which beacon servers in each AS generate Path Construction Beacons (PCBs). As PCBs propagate towards ISD cores, each AS appends cryptographically signed hop information, creating validated path segments. Segments are classified as up-segments (from the source AS towards its ISD core), core-segments (between ISD cores), and down-segments (from a remote ISD core to the destination AS). By combining up, core, and down segments, endpoints can assemble multiple valid end-to-end paths. Crucially, these segments have a finite lifetime and must be refreshed by new beacons. This process of expiration and renewal is a core design principle that makes the set of available paths inherently dynamic~\cite{krahenbuhl2024glids}. 
    
    The data plane uses packet-carried forwarding state (PCFS) as the complete inter-domain path (including ingress/egress interfaces) is embedded in every packet's header. This enables stateless forwarding between ASes and removes the need for large inter-domain forwarding tables.
    
    For multipath transport, SCION's architecture presents both opportunities (many topologically diverse, policy-compliant paths) and challenges (a potentially large and dynamic path set with variable performance). Two notions of stability are relevant for transport protocols: (1) \emph{control-plane stability}—the persistence of discoverable paths over time—and (2) \emph{data-plane stability}—the consistency of performance metrics (latency, loss, jitter) for a given discovered path. The interaction between these layers determines the effectiveness of multipath transport in SCION. 
    
    For a comprehensive technical specification, see the SCION book~\cite{chuat2022complete} and prior architectural analyses~\cite{zhang2011scion}.

\subsection{Related Work}~\label{subsec::relatedwork}
    Kwon et al.~\cite{kwon2020scionlab} introduced SCIONLab as a global research network for testing the SCION architecture, enabling the attachment of user ASes and facilitating real-world experimentation with path-aware networking. Building on this, Kr"{a}henb"{u}hl et al.~\cite{krahenbuhl2021deployment} evaluated the deployment and scalability of SCION's inter-domain multipath routing infrastructure, analyzing path construction algorithms and their ability to achieve near-optimal path diversity. More recently, van Rijswijk-Deij et al.~\cite{vanrijswijk2023sciontestbed} presented a SCION performance benchmarking suite deployed across multiple testbed sites, covering latency, throughput, and other metrics. While their methodology aligns with ours, their evaluation primarily targets production-grade metrics with a focus on SCION's deployment maturity. In contrast, our study emphasizes the dynamism of SCION's multipath capabilities, examining path churn rates and performance asymmetries under SCIONLab’s current non-production conditions. Other measurements in SCIONLab include a study by De et al.~\cite{battipaglia2023evaluation}, who investigated the performance of user-driven path control from a usability perspective.
    
    Further research has focused on path analysis, optimization, and stability in SCION networks. Van Veen~\cite{van2024achieving} developed methods for application-level requirement-based path selection, allowing endpoints to choose paths based on specific criteria like low latency or high bandwidth while considering policy constraints. Gessner~\cite{gessner2021leveraging} explored application-layer path awareness to enhance multipath connectivity, proposing mechanisms to dynamically select and combine paths for improved performance. Koole and Pawlus~\cite{koole2023comparative} provided a comparative analysis of routing policies in BGP and SCION, highlighting how SCION's policy-driven approach affects path stability, diversity, and overall network resilience.
    
    Several works have investigated multipath transport protocols tailored to SCION's path-aware environment. John et al.~\cite{john2023dmtp} proposed DMTP, a deadline-aware multipath transport protocol that leverages SCION's multiple paths to optimize for latency-sensitive applications. Similarly, Gartner et al.~\cite{gartner2023hercules} introduced Hercules, a high-speed bulk-transfer system that exploits path diversity for enhanced throughput.
    Our work provides the empirical characterization of path dynamics, such as churn, stability, and asymmetry, that is essential for the design and realistic evaluation of such advanced transport systems.

\section{Measurement Methodology}~\label{sec::methodology}
    To systematically characterize the SCIONLab environment, we designed and deployed a custom measurement suite on three identical Google Cloud Virtual Machines. This section details our methodology, beginning with the experimental setup, followed by the specific scripts in our measurement suite, and concluding with our data analysis pipeline. The complete framework, including all scripts and analysis tools, is available in our public repository to ensure full reproducibility~\cite{github-scionlab-data}.

    \subsection{Experimental Setup}
    Our three measurement nodes were provisioned as SCIONLab ASes. While they were attached to geographically diverse attachment points to ensure diverse end-to-end SCION paths, specifically, ISD 19 (Europe), ISD 17 (Switzerland), and ISD 18 (US), the virtual machines themselves were all hosted in the \texttt{us-central1-a} Google Cloud region. This co-location was a deliberate methodological choice. By originating all measurements from a single, well-provisioned network location, we minimized confounding performance variables from the underlying cloud provider's wide-area network. This ensures that the observed latency, loss, and jitter are more directly attributable to the dynamics of the SCION paths being measured, rather than artifacts of the legacy IP transit connecting the measurement nodes to the SCION infrastructure.

\begin{figure}[htbp]
      \centering
      \includegraphics[width=\textwidth]{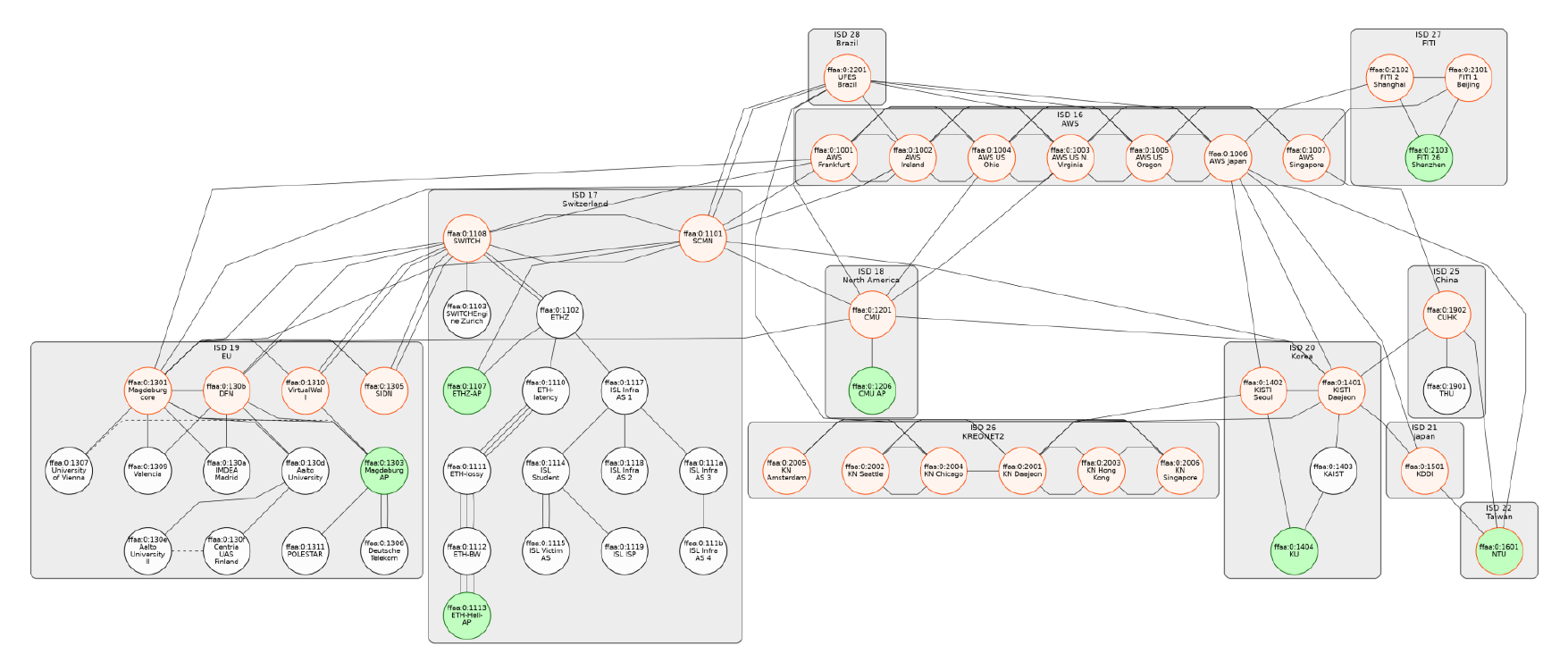}
      \caption{The global topology of the SCIONLab testbed. This figure provides an overview of the network structure, though we note that not all Autonomous Systems (ASes) shown were reachable during our measurement campaign. Original image sourced from SCIONLab.org.}
      \label{fig:scionlab-topology}
    \end{figure}

    These measurement ASes were configured to conduct a full mesh of measurements between each other, meaning AS-1 probes AS-2 and AS-3, AS-2 probes AS-1 and AS-3, and so forth. Figure~\ref{fig:scionlab-topology} provides an overview of the SCIONLab network topology at the time of our study. However, we note that not all Autonomous Systems depicted were reachable during our measurement campaign.
    
    Our measurement campaign ran continuously for four weeks, from July 11th to August 11th, 2025. After addressing initial deployment instabilities, we focused our analysis on the stable period from July 16th to August 11th to ensure the integrity of our findings. Measurements between our three ASes were executed by a cron job in 30-minute intervals, with the raw data being archived for later analysis.

\subsection{Measurement Suite}
    Our measurement suite consists of a pipeline of Python scripts designed to probe the SCION control and data planes at each interval.
    
    \begin{description}
        \item[pathdiscover] Utilizes the \texttt{scion showpaths} command to enumerate all available end-to-end paths to our target ASes.

        \item[comparer] Compares the newly discovered path set against the set from the previous interval to log path appearance and disappearance events, forming the basis of our churn analysis.
        
        \item[prober] Probes data-plane performance by running \texttt{scion ping} over a random sample of up to 15 available paths to each target. We chose this sample size to balance the need for a statistically representative snapshot against the risk of creating excessive network load on the shared testbed.
        
        \item[mp-prober] Assesses multipath performance by initiating up to three simultaneous \texttt{scion ping} instances over a distinct subset of the discovered paths. Our goal was to measure the impact of concurrency on per-path performance, not to maximize aggregate throughput, hence the choice of a small, three-path set (e.g., primary, secondary, tertiary). To ensure a fair comparison, the paths used in these multipath tests were a subset of those probed individually by the \textbf{prober} script in the same interval.
        
        \item[bw-alldiscover \& bw-multipath] Employs the \texttt{scion-bwtestclient} to measure throughput. We limited these tests to a maximum of two paths per destination to minimize the intrusive impact of high-volume traffic on the non-production SCIONLab network.
    \end{description}
    All scripts output their raw data into dated directories for later use.

\subsection{Analysis Suite}
    The analysis suite processes the collected JSON data on a per-AS basis. The tools extract key metrics, including average Round-Trip Time (RTT), jitter, packet loss, and path churn rates, and generate the visualizations presented in Section~\ref{sec::result}. To maintain clarity, the analysis concentrates on the most representative results that illustrate the core findings of our study. The complete raw dataset and analysis scripts are publicly available in our repository to ensure full reproducibility of our work~\cite{github-scionlab-data}.

\section{Results}~\label{sec::result}
    This section presents the results of our measurement study. A preliminary analysis of the data from all three of our geographically distributed vantage points revealed highly consistent trends in path dynamics and performance characteristics. Therefore, to ensure clarity and provide a detailed, uncluttered presentation, the figures and primary analysis in this paper focus on the dataset from a single vantage point as a representative case study. The corroborating data from our other vantage points is included in our publicly available data repository~\cite{github-scionlab-data}.

\subsection{Path Availability and Churn}
\begin{figure}[h!]
      \centering
      \includegraphics[width=\textwidth]{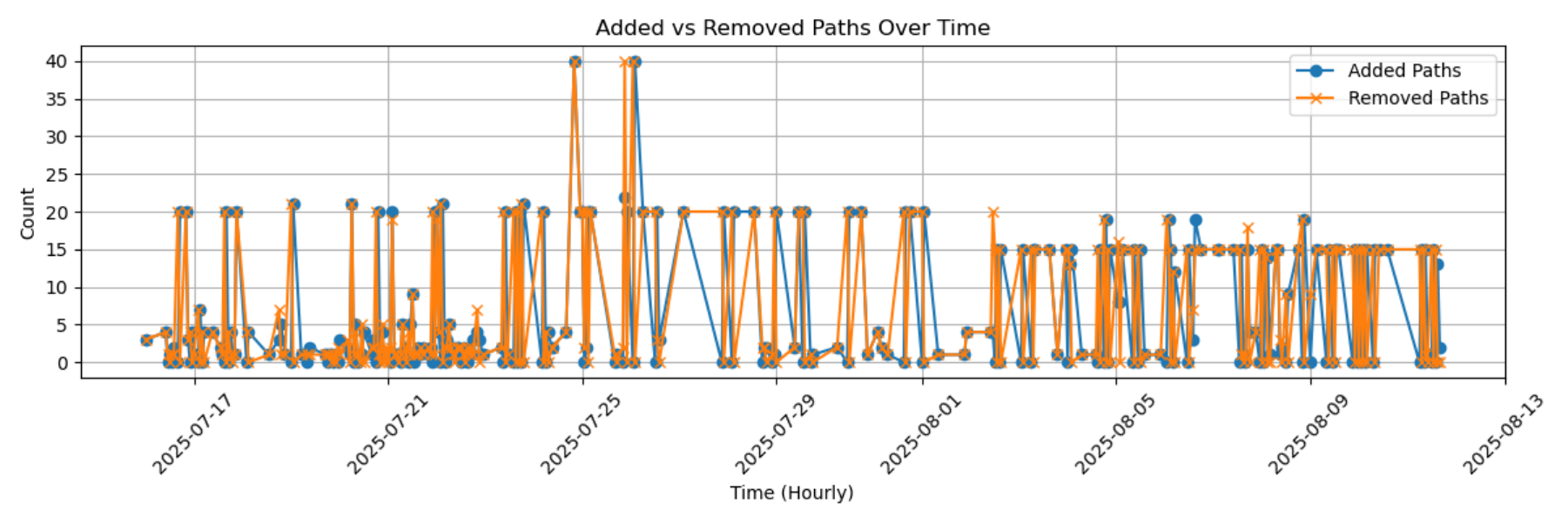}
      \caption{Path churn over the measurement period, showing the number of paths added and removed between the measurement AS and its destinations during each 30-minute interval.}
      \label{fig:churn-events}
    \end{figure}

\begin{figure}[h!]
      \centering
      \includegraphics[width=\textwidth]{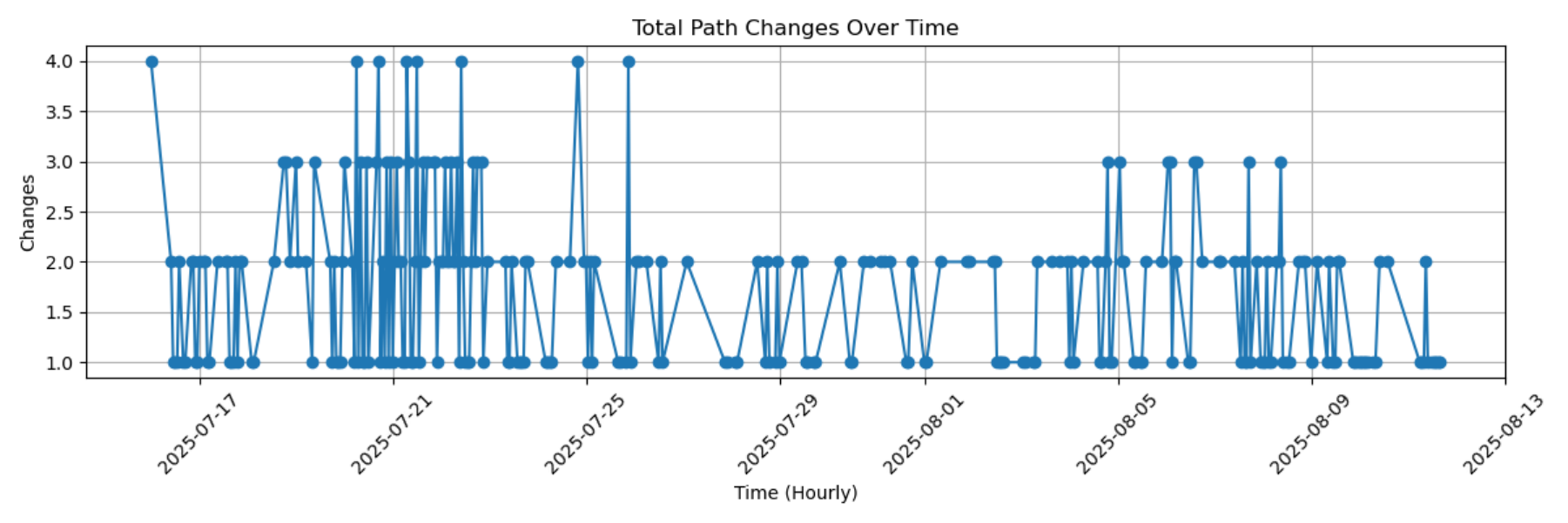}
      \caption{Frequency of path change events over time. Each data point signifies an interval where the set of available paths changed. Intervals with no changes were excluded.}
      \label{fig:change-events}
    \end{figure}

    Our measurements reveal a highly dynamic path environment in SCIONLab, characterized by significant and constant path churn. The set of available end-to-end paths was in constant flux, with paths frequently being added and removed throughout the measurement period, as illustrated in Figure~\ref{fig:churn-events}. This instability is further underscored by Figure~\ref{fig:change-events}, which shows that path change events occurred in the majority of our 30-minute observation intervals.

    To quantify this churn, we tracked changes between our measurement AS and its two target destinations. For destination \texttt{17-ffaa:1:11e4}, we recorded 161 change events over 1281 comparisons, with 557 paths added and 577 removed. This resulted in an average path lifetime of approximately 15.7 hours. The churn was even more pronounced for destination \texttt{18-ffaa:1:11e5}, which experienced 301 change events with 1254 paths added and 1259 removed, yielding a shorter average path lifetime of just \textbf{8.6 hours}.

\begin{figure}[h!]
      \centering
      \includegraphics[width=\textwidth]{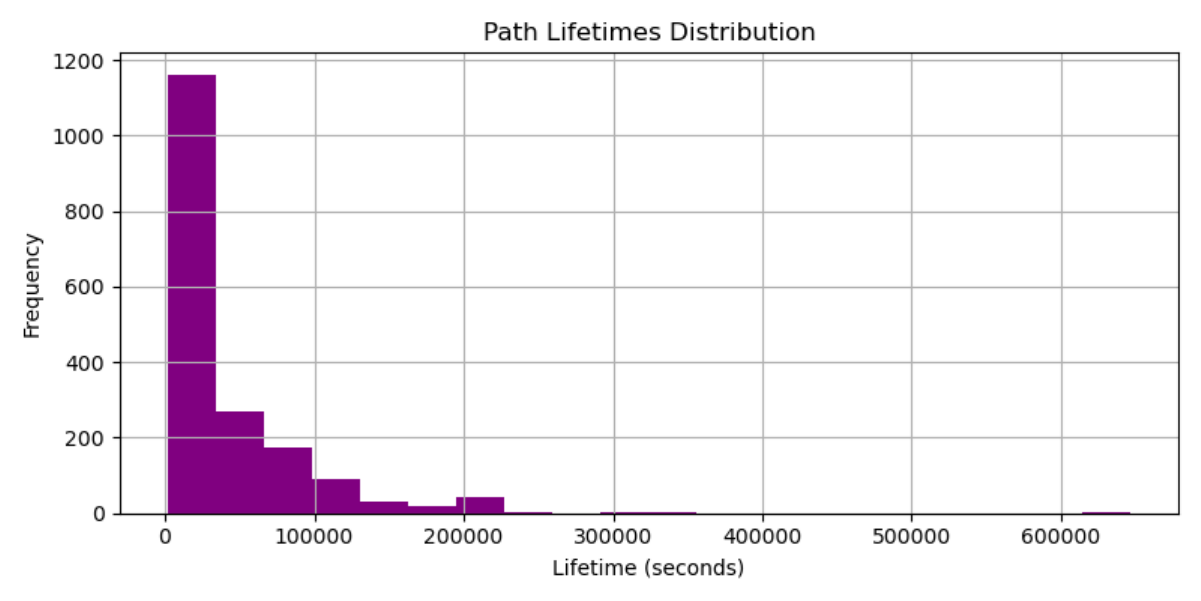}
      \caption{Distribution of path lifetimes. The histogram clearly shows that the majority of paths discovered during the experiment are ephemeral and short-lived.}
      \label{fig:lifetime-dist}
    \end{figure}

    A direct consequence of this high churn rate is that paths are overwhelmingly ephemeral, as confirmed by the path lifetime distribution in Figure~\ref{fig:lifetime-dist}. The vast majority of paths lasted for less than 100,000 seconds (approx. 27 hours). The ephemeral nature of these paths challenges the assumption of path stability, a critical consideration for transport protocols like MPQUIC. Such protocols must be designed to be resilient to frequent and rapid changes in the set of available paths.
    
\subsection{Latency and Loss Analysis}
    To assess data-plane performance, we probed the latency and loss characteristics across a random sample of up to 15 available paths for each destination. The aggregate results revealed performance trade-offs between the two targets:
    \begin{itemize}
        \item Paths to \texttt{AS 17-ffaa:1:11e4} had a higher average RTT of 470.96~ms but a lower packet loss rate of 0.36\%.
        \item In contrast, paths to \texttt{AS 18-ffaa:1:11e5} offered lower average latency (RTT 404.97~ms) but exhibited higher packet loss (0.53\%).
    \end{itemize}

\begin{figure}[h!]
      \centering
      \includegraphics[width=\textwidth]{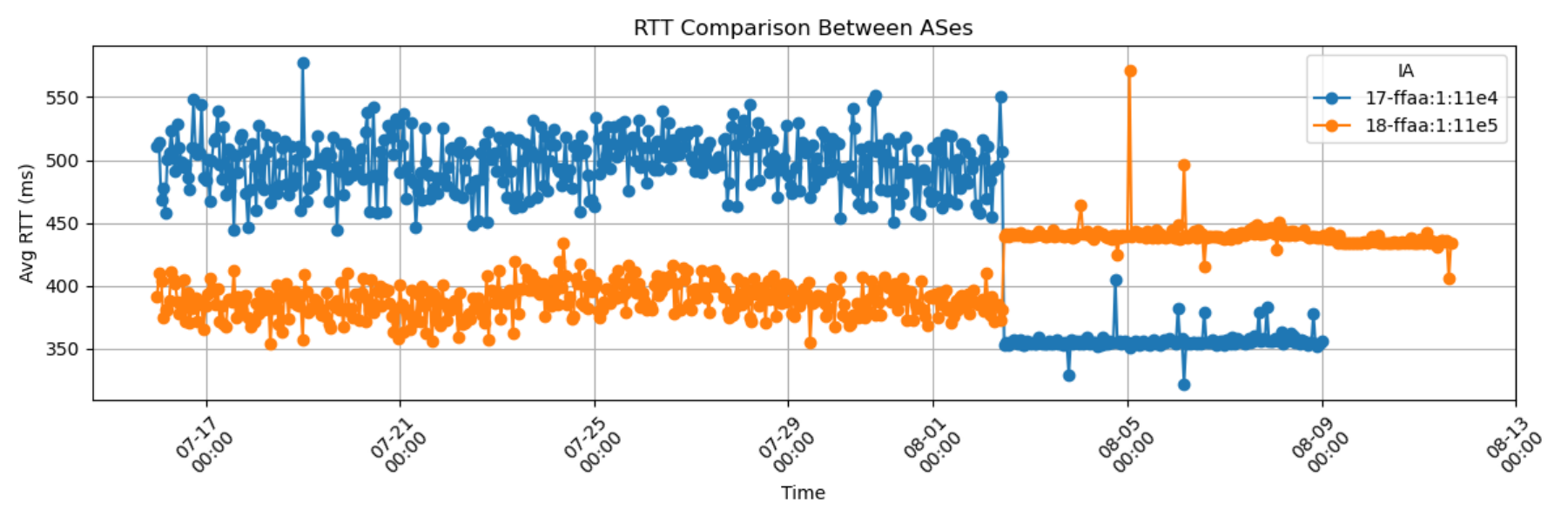}
      \caption{Average Round-Trip Time (RTT) to both target ASes over the measurement period. A distinct stabilization and shift in latency for both destinations is visible starting around August 2nd.}
      \label{fig:rtt-over-time}
    \end{figure}

\begin{figure}[h!]
      \centering
      \includegraphics[width=\textwidth]{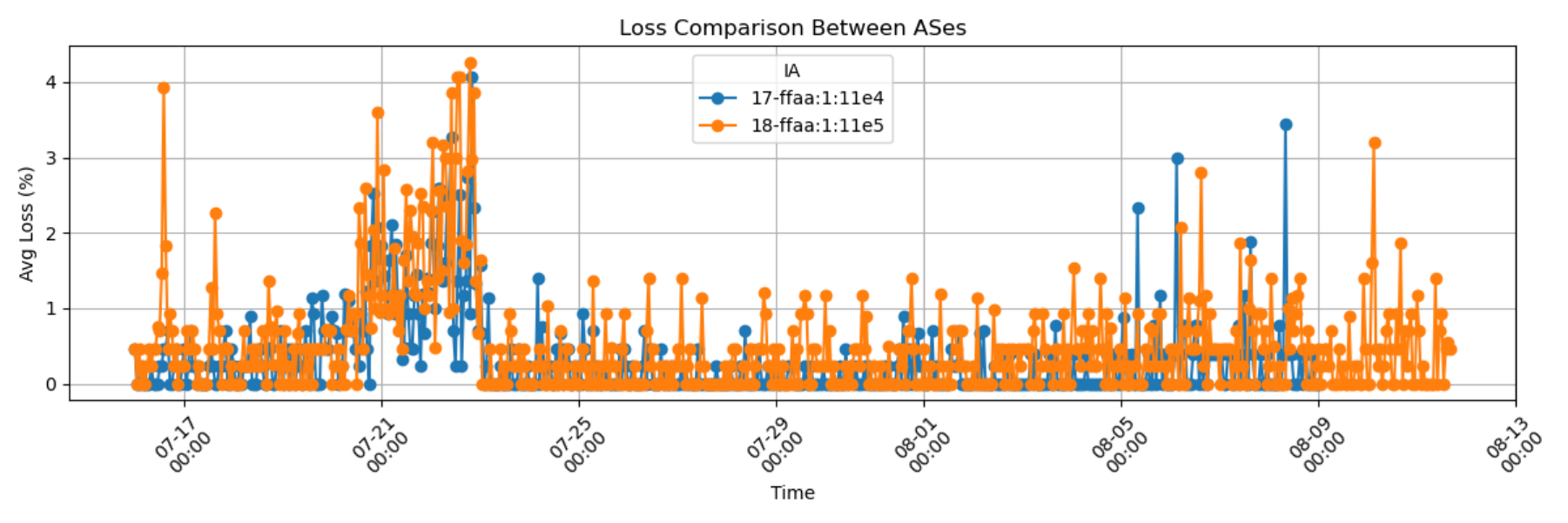}
      \caption{Average packet loss to both target ASes over the same period. In contrast to the RTT measurements, the packet loss behavior does not show a corresponding stabilization event.}
      \label{fig:loss-over-time}
    \end{figure}

    Beyond these averages, the time-series data exposed a notable phenomenon. As shown in Figure~\ref{fig:rtt-over-time}, a distinct stabilization event occurred around August 2nd, after which the RTT values for both destinations became significantly less volatile. This effect was observed across all our measurement ASes, suggesting a SCIONLab-wide event. Interestingly, this dramatic stabilization in latency did not correlate with reliability; Figure~\ref{fig:loss-over-time} shows that packet loss patterns remained erratic. Our hypothesis is that this RTT stabilization was caused by the disappearance of multiple unstable, high-latency paths from the network topology around that time.

\subsection{Bandwidth Performance Asymmetry}
\begin{figure}[h!]
      \centering
      \includegraphics[width=\textwidth]{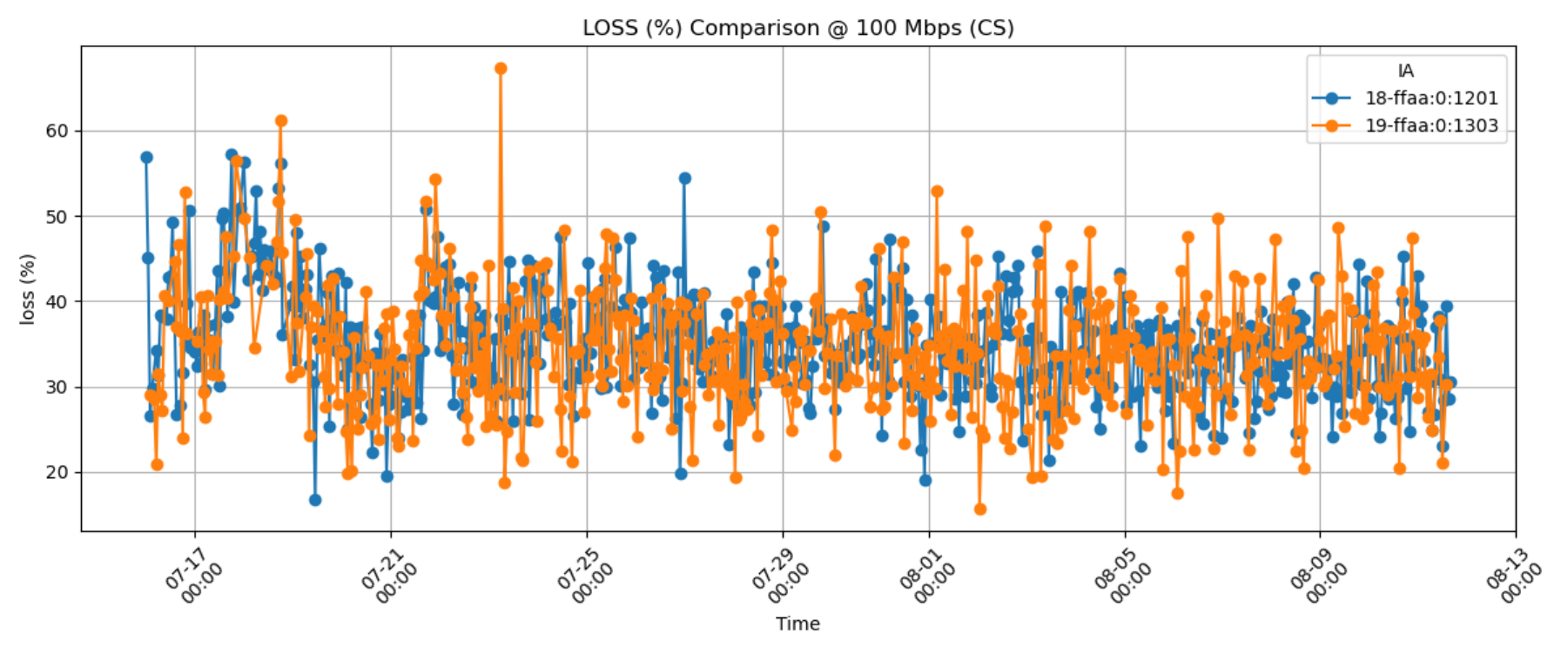}
      \caption{Packet loss in the upstream (client-to-server, $C \rightarrow S$) direction at the 100 Mbps bandwidth tier. The plot shows persistently high and volatile packet loss, indicating a significant performance bottleneck.}
      \label{fig:bw-loss-cs}
    \end{figure}

    Our bandwidth tests revealed a significant performance asymmetry between the downstream and upstream directions, which became particularly severe at higher throughput tiers. While the 100~Mbps downstream tests performed well, the upstream tests exposed a major bottleneck:
    \begin{itemize}
        \item \textbf{Downstream ($S \rightarrow C$)}: Server-to-client traffic maintained a reasonably high throughput, averaging 87.09~Mbps.
        \item \textbf{Upstream ($C \rightarrow S$)}: In sharp contrast, client-to-server traffic suffered dramatically, with average bandwidth dropping to 65.4~Mbps, driven by extremely high packet loss that averaged \textbf{34.6\%}, as shown in Figure~\ref{fig:bw-loss-cs}.
    \end{itemize}
    This persistent and volatile upstream packet loss strongly suggests the presence of significant bottlenecks or restrictive routing policies within the SCIONLab testbed infrastructure.

\subsection{Single-Path vs. Multipath Performance Trade-offs}
    To understand the impact of concurrent path usage, we compared the performance of individual paths when used in isolation (Single-Path, SP) versus simultaneously in a group (Multipath, MP). Our analysis revealed a critical and counter-intuitive trade-off: using multiple paths at once degrades the performance of each constituent path.
    
\begin{figure}[h!]
        \centering
        \begin{subfigure}{\textwidth}
            \centering
            \includegraphics[width=0.9\linewidth]{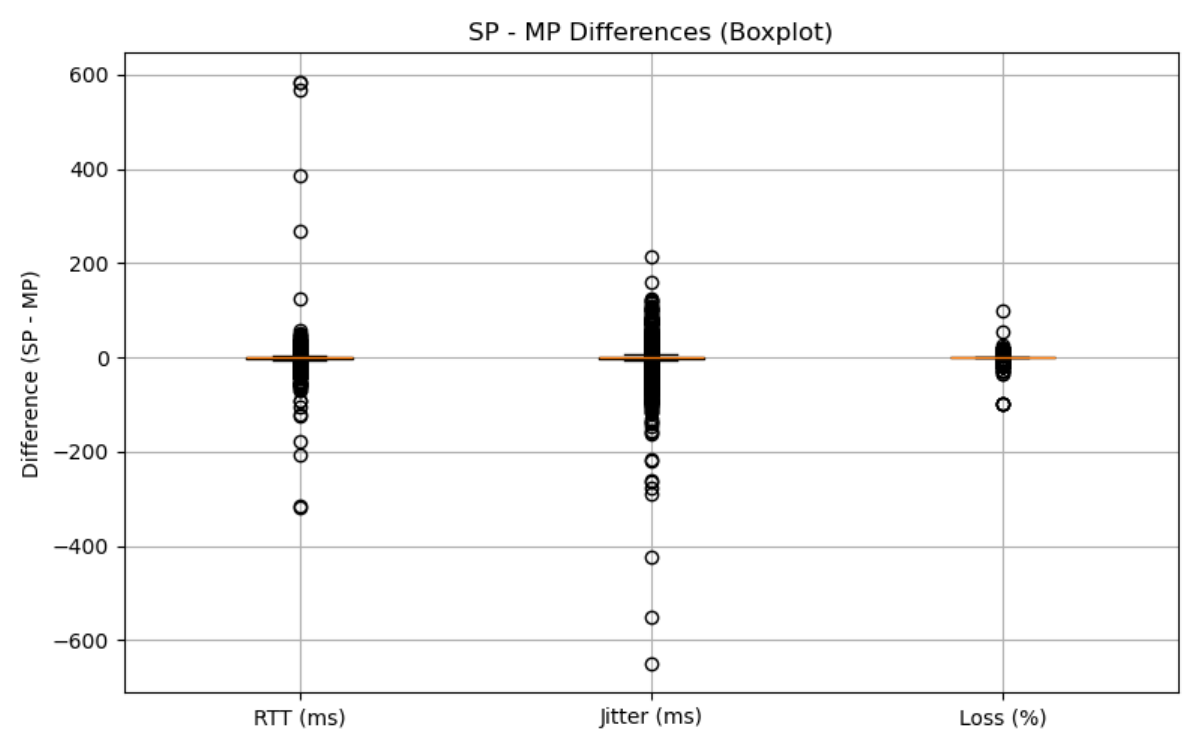}
            \subcaption{Performance differences for latency metrics ($SP - MP$). Negative medians indicate that individual paths performed worse (e.g., had higher latency or loss) when used concurrently.}
            \label{fig:sp-mp-latency}
        \end{subfigure}
        \vfill 
        \begin{subfigure}{\textwidth}
            \centering
            \includegraphics[width=0.9\linewidth]{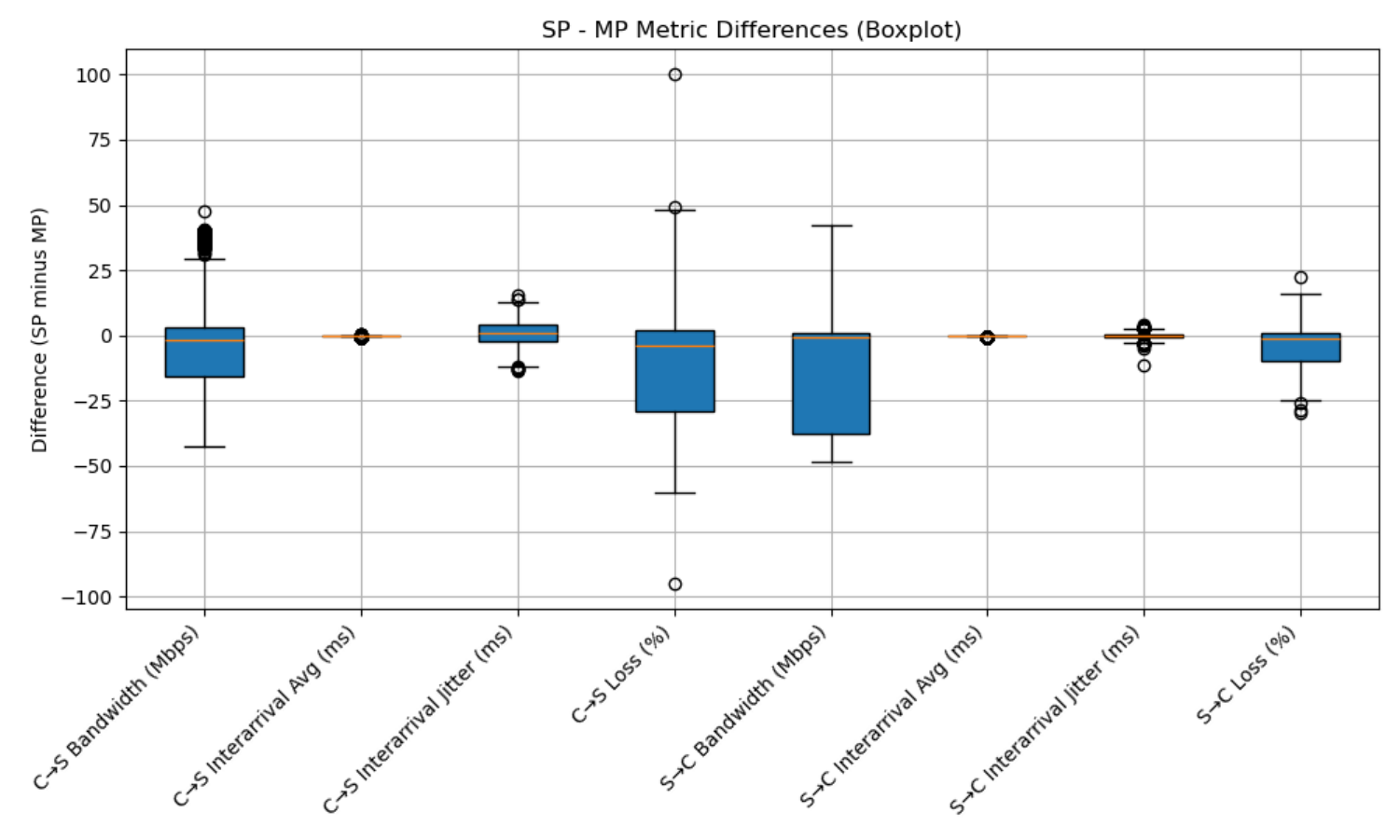}
            \subcaption{Performance differences for bandwidth metrics ($SP - MP$). While per-path performance is mixed, the total aggregate throughput in MP mode was higher.}
            \label{fig:sp-mp-bandwidth}
        \end{subfigure}
        \caption{Per-path performance comparison between Single-Path (SP) and Multipath (MP) measurements. Concurrent usage degrades per-path latency metrics (a) but provides a higher total aggregate throughput (b).}
        \label{fig:sp-mp-combined}
    \end{figure}

    As Figure~\ref{fig:sp-mp-combined}a illustrates, paths in an MP context consistently showed slightly worse latency-sensitive metrics, with higher average RTT, increased jitter, and greater packet loss compared to their SP counterparts. This suggests that concurrent flows introduced contention, either at the end-hosts or at shared bottleneck links within the SCIONLab infrastructure.

    However, this per-path degradation is the cost of achieving higher aggregate throughput. While the per-path bandwidth measurements under MP conditions were mixed (Figure~\ref{fig:sp-mp-combined}b), the combined throughput of the multipath session consistently outperformed any single path. This clarifies a fundamental trade-off for multipath transport in this environment: applications sacrifice a minor degree of per-path quality for a significant gain in total available bandwidth. This implies that advanced schedulers for protocols like MPQUIC should not treat all paths equally. They could strategically route latency-sensitive traffic over the best single path while leveraging multiple paths for bulk data transfer to maximize throughput.

\subsection{Path Discrepancy: Asymmetric Path Discovery}
    Our study revealed the phenomenon of \textit{path discrepancy}, where path availability between two endpoints is not symmetric. We observed, for instance, a case where AS-3 could discover 20 viable paths to AS-2, while AS-2 could only find 18 paths in the reverse direction to AS-3. This is not a transient failure but an expected consequence of SCION's design, where ASes have autonomy over which path segments they accept and advertise. This has direct implications for protocols like MPQUIC, which may assume that if a sender can use a path, the receiver can use it for return traffic. Our findings show this assumption is invalid in SCION. Accounting for this discrepancy is crucial for applications that benefit from bidirectional consistency.

    \section{Discussion}~\label{sec::discussion}
    Our longitudinal measurements on the SCIONLab testbed provide a crucial, empirically-grounded characterization of the dynamics in a deployed path-aware network. While SCIONLab is not a production-grade deployment, its nature as an overlay on the public internet offers a realistic glimpse into the challenges that protocols will face in early-stage or hybrid SCION environments. Our findings on path churn, asymmetry, and multipath performance have direct and significant implications for the design of next-generation transport protocols like MPQUIC.

    A primary finding of this study is the high rate of control-plane path churn, with average path lifetimes as short as 8.6 hours. This environment is far more dynamic than the BGP-based internet, presenting a fundamental challenge to multipath protocols. Schedulers for MPQUIC cannot treat the set of available paths as static; they must be designed with the explicit assumption that paths can and will disappear frequently. This necessitates mechanisms for continuous, low-overhead path probing and validation to quickly adapt to failures. For deadline-sensitive protocols like DMTP~\cite{john2023dmtp}, such churn is not merely a performance issue but a critical reliability concern. A scheduler that assigns a time-critical flow to a path that subsequently vanishes must have a robust and rapid failover strategy. Our findings suggest that path stability should be a primary metric for path selection, perhaps even more so than latency or bandwidth for certain applications.

    Our work also reveals that path availability is frequently asymmetric, a phenomenon we term path discrepancy. This is not a network fault but an expected consequence of SCION’s design, which empowers individual ASes with independent path-filtering policies. This has profound implications for protocols like MPQUIC, as many scheduling and congestion control algorithms implicitly assume path symmetry for efficient operation (e.g., using forward-path RTT to time reverse-path acknowledgements). Our findings show this assumption is unsafe in SCION. This control-plane asymmetry is further compounded by the severe data-plane bandwidth asymmetry we observed, with upstream packet loss reaching over 34\%. High-speed transfer systems like Hercules~\cite{gartner2023hercules} must explicitly measure and account for this directional disparity to achieve optimal performance.
    
    Furthermore, our results illuminate a critical performance trade-off inherent in multipath transport. We found that using multiple paths concurrently can degrade the latency and loss characteristics of each individual path, likely due to end-host or network-level resource contention. However, this minor per-path degradation is the price for a significant gain in aggregate throughput. This finding offers clear guidance for scheduler design: a sophisticated MPQUIC scheduler should not treat all paths as equals. It could strategically isolate a single, high-quality path for latency-sensitive streams (e.g., control messages, VoIP) while distributing high-volume, throughput-oriented traffic across the remaining paths, maximizing total bandwidth.

    \paragraph{Limitations and Future Directions} We must emphasize that our findings are derived from the SCIONLab testbed, an experimental network overlaid on the public IP internet. Therefore, it is possible that some of the observed dynamics, particularly performance artifacts like high upstream loss, could be influenced by the underlying VPN tunnels or cloud provider infrastructure rather than being inherent to SCION itself. Consequently, these results should be interpreted as a foundational characterization of a developmental path-aware network, not as the definitive performance of a future, scaled-out SCION internet. The next essential step is to deploy this measurement framework in production-grade SCION environments, such as SCIERA. This will allow for a comparative analysis to validate our findings and disentangle the fundamental properties of SCION's architecture from the implementation-specific artifacts of the testbed.

    \section{Conclusion}~\label{sec::conclusion}
    In this paper, we presented a longitudinal measurement study of the SCIONLab testbed to characterize the real-world dynamics of a deployed path-aware network. Our findings reveal a highly volatile environment governed by three key properties: 1) significant control-plane churn, with average path lifetimes as short as 8.6 hours; 2) path discrepancy, an inherent asymmetry in path availability caused by routing policies; and 3) a fundamental performance trade-off where concurrent multipath usage improves aggregate throughput at the cost of degrading per-path latency and reliability. Taken together, these results provide data-driven evidence that next-generation multipath protocols like MPQUIC must be designed to handle high churn and inherent asymmetry, challenging core assumptions of path stability and symmetry derived from the traditional Internet.

\section*{CRediT authorship contribution statement}
\textbf{Lars Herschbach}: Conceptualization, Methodology, Software, Data curation, Formal analysis, Visualization, Writing -- original draft, Writing -- review \& editing.\newline
\textbf{Damien Rossi}: Software, Resources, Validation. \newline
\textbf{Sina Keshvadi}: Supervision, Methodology, Validation, Writing -- review \& editing. \newline

\section*{Declaration of competing interest}
The authors declare that they have no known competing financial interests or personal relationships that could have appeared to influence the work reported in this paper.

\section*{Funding}
This research received no specific grant from funding agencies in the public, commercial, or not-for-profit sectors. Lars Herschbach and Damien Rossi were part of the Mitacs GRI internship Program which funded their stay in Canada. No Mitacs funding was intended for, or used in, the creation of this paper.
We would also like to thank Google for providing us with Google cloud credits making the persistent measurements using their infrastructure possible.

\section*{Data availability}
The datasets and analysis scripts generated during and/or analyzed during the current study are available in the \href{https://github.com/Keshvadi/mpquic-on-scion-ipc/tree/main}{GitHub repository}~\cite{github-scionlab-data}.

\section*{AI use statement}
The authors used OpenAI’s ChatGPT to assist in language editing and structuring of the manuscript as well as in the creation of the measurement and analysis scripts. The authors reviewed, verified, and edited the content, and take full responsibility for its content.

\bibliography{main}

\begin{thebibliography}{10}
\expandafter\ifx\csname url\endcsname\relax
  \def\url#1{\texttt{#1}}\fi
\expandafter\ifx\csname urlprefix\endcsname\relax\def\urlprefix{URL }\fi
\expandafter\ifx\csname href\endcsname\relax
  \def\href#1#2{#2} \def\path#1{#1}\fi

\bibitem{giaconia2025crunching}
A.~Giaconia, M.~Tran, L.~Vanbever, S.~Vissicchio, Is crunching public data the
  right approach to detect bgp hijacks?, arXiv preprint arXiv:2507.20434
  (2025).

\bibitem{butler2009survey}
K.~Butler, T.~R. Farley, P.~McDaniel, J.~Rexford, A survey of bgp security
  issues and solutions, Proceedings of the IEEE 98~(1) (2009) 100--122.

\bibitem{clark2018designing}
D.~D. Clark, Designing an internet, MIT Press, 2018.

\bibitem{zhang2011scion}
X.~Zhang, H.-C. Hsiao, G.~Hasker, H.~Chan, A.~Perrig, D.~G. Andersen, Scion:
  Scalability, control, and isolation on next-generation networks, in: 2011
  IEEE Symposium on Security and Privacy, IEEE, 2011, pp. 212--227.

\bibitem{anapaya2025}
A.~S. AG, Secure, resilient, and controlled connectivity with scion,
  \url{https://www.anapaya.net/}, accessed: 2025-08-11 (2025).

\bibitem{rfc8684}
A.~Ford, C.~Raiciu, M.~J. Handley, O.~Bonaventure, C.~Paasch,
  \href{https://www.rfc-editor.org/info/rfc8684}{{TCP Extensions for Multipath
  Operation with Multiple Addresses}}, RFC 8684 (Mar. 2020).
\newblock \href {https://doi.org/10.17487/RFC8684}
  {\path{doi:10.17487/RFC8684}}.
\newline\urlprefix\url{https://www.rfc-editor.org/info/rfc8684}

\bibitem{de2017multipath}
Q.~De~Coninck, O.~Bonaventure, Multipath quic: Design and evaluation, in:
  Proceedings of the 13th international conference on emerging networking
  experiments and technologies, 2017, pp. 160--166.

\bibitem{ietf-quic-multipath-15}
Y.~Liu, Y.~Ma, Q.~D. Coninck, O.~Bonaventure, C.~Huitema, M.~Kühlewind,
  \href{https://datatracker.ietf.org/doc/draft-ietf-quic-multipath/15/}{{Multipath
  Extension for QUIC}}, Internet-Draft draft-ietf-quic-multipath-15, Internet
  Engineering Task Force, work in Progress (Jul. 2025).
\newline\urlprefix\url{https://datatracker.ietf.org/doc/draft-ietf-quic-multipath/15/}

\bibitem{krahenbuhl2024glids}
C.~Kr{\"a}henb{\"u}hl, S.~Tabaeiaghdaei, S.~Scherrer, M.~Frei, A.~Perrig,
  Glids: A global latency information dissemination system, arXiv preprint
  arXiv:2405.04319 (2024).

\bibitem{krahenbuhl2021deployment}
C.~Kr{\"a}henb{\"u}hl, S.~Tabaeiaghdaei, C.~Gloor, J.~Kwon, A.~Perrig,
  D.~Hausheer, D.~Roos, Deployment and scalability of an inter-domain
  multi-path routing infrastructure, in: Proceedings of the 17th International
  Conference on emerging Networking EXperiments and Technologies, 2021, pp.
  126--140.

\bibitem{kwon2020scionlab}
J.~Kwon, J.~A. Garc{\'\i}a-Pardo, M.~Legner, F.~Wirz, M.~Frei, D.~Hausheer,
  A.~Perrig, Scionlab: A next-generation internet testbed, in: 2020 IEEE 28th
  International Conference on Network Protocols (ICNP), IEEE, 2020, pp. 1--12.

\bibitem{chuat2022complete}
L.~Chuat, M.~Legner, D.~Basin, D.~Hausheer, S.~Hitz, P.~M{\"u}ller, A.~Perrig,
  The complete guide to scion, Information Security and Cryptography (2022).

\bibitem{vanrijswijk2023sciontestbed}
L.~van Rijswijk-Deij, C.~de~Laat,
  \href{https://delaat.net/sc/sc23/indis/S4-4-sc23_presenter_leo.pdf}{{SCION
  Performance Testbed Evaluation}}, SC23 INDIS Presentation, accessed:
  2025-07-23 (2023).
\newline\urlprefix\url{https://delaat.net/sc/sc23/indis/S4-4-sc23_presenter_leo.pdf}

\bibitem{battipaglia2023evaluation}
A.~Battipaglia, L.~Boldrini, R.~Koning, P.~Grosso, Evaluation of scion for
  user-driven path control: A usability study, in: Proceedings of the SC'23
  Workshops of The International Conference on High Performance Computing,
  Network, Storage, and Analysis, 2023, pp. 785--794.

\bibitem{van2024achieving}
A.~van Veen, H.~Vranken, R.~Koning, C.~Schutijser, E.~Poll, Achieving
  application-level requirement-based path selection within scion (2024).

\bibitem{gessner2021leveraging}
J.~Gessner, Leveraging application layer path-awareness with scion, Master's
  thesis, ETH Zurich (2021).

\bibitem{koole2023comparative}
K.~Koole, M.~Pawlus, A comparative analysis of routing policies in bgp and
  scion (2023).

\bibitem{john2023dmtp}
T.~John, A.~Perrig, D.~Hausheer, Dmtp: Deadline-aware multipath transport
  protocol, in: 2023 IFIP Networking Conference (IFIP Networking), IEEE, 2023,
  pp. 1--9.

\bibitem{gartner2023hercules}
M.~Gartner, J.-P. Smith, M.~Frei, F.~Wirz, C.~Neukom, D.~Hausheer, A.~Perrig,
  Hercules: High-speed bulk-transfer over scion, in: 2023 IFIP Networking
  Conference (IFIP Networking), IEEE, 2023, pp. 1--9.

\bibitem{github-scionlab-data}
L.~Herschbacha, D.~Rossi, S.~Keshvadi, mpquic-on-scion-ipc,
  \url{https://github.com/Keshvadi/mpquic-on-scion-ipc} (2025).

\end{thebibliography}

\end{document}